\documentclass[fleqn,twoside]{article}
\usepackage{espcrc2}

\usepackage{graphicx}

\newcommand{\noi}{\noindent}
\newcommand{\eq}{\begin{equation}}
\newcommand{\en}{\end{equation}}
\newcommand{\eqa}{\begin{eqnarray}}
\newcommand{\ena}{\end{eqnarray}}

\newcommand{\oeta}{{\overline \eta}}

\newcommand{\opsi}{{\overline \psi}}

\newcommand{\ozeta}{{\overline \zeta}}

\newcommand{\cam}{{\cal M}}
\newcommand{\cao}{{\cal O}}
\newcommand{\capbig}{{\cal P}}

\newcommand{\hatJ}{{\widehat J}}
\newcommand{\hatH}{{\widehat H}}

\newcommand{\hatN}{{\widehat N}}

\newcommand{\hatT}{{\hat T}}

\newcommand{\hatV}{{\widehat V}}

\newcommand{\bfB}{{\bf B}}
\newcommand{\bfC}{{\bf C}}

\newcommand{\tr}{\mbox{Tr}\,}

\newcommand{\vx}{{\vec x}}
\newcommand{\vy}{{\vec y}}

\newcommand{\AmS}{{\protect\the\textfont2
  A\kern-.1667em\lower.5ex\hbox{M}\kern-.125emS}}

\title{Currents, chemical potential and boundary conditions in lattice QCD
}

\author{V.K. Mitrjushkin\address[JINR]{Joint Institute for Nuclear Research,
         141980 Dubna, Russia}
         \address[ITEP]{Institute of Theoretical and Experimental Physics,
                    Moscow, Russia}
        \thanks{Talk given at {\it LATTICE 2002} Int.Symp. (June 24--29, 2002,
        Cambridge, Massachusetts, USA)}
        \thanks{Supported by the grant INTAS--00--00111 and RFBR grant
                02--02--17308.  }
}

\begin{document}

\begin{abstract}
A connection between the operator fermionic currents $~\hatJ~$ and
corresponding `Grassmannian' currents $J$ in the functional integral
formalism is studied. The QCD action with non--zero chemical potential
$\mu$ is derived. A connection between the fermionic Fock space and
boundary conditions along the forth direction is discussed.

\vspace{1pc}
\end{abstract}

\maketitle

\section{Introduction}

One needs transfer matrix (TM) formalism :

\noi {\bf a)} ~to define boundary conditions (b.c.'s) for $\psi,\opsi_x$
and $U_{x\mu}$ in the functional integrals, in particular,   along the
forth (`time') direction ;

\noi {\bf b)} ~to establish connection between correlators of currents
$\hatJ$ and corresponding masses.

In Wilson approach the average of any functional
$~\cao(U;\psi;\opsi)~$ is \cite{wil1}

\eq
\langle \cao\rangle_W = \frac{1}{Z_W}\int\! [dU][d\psi d\opsi]
~\cao(U;\psi;\opsi)\cdot e^{-S_W}
            \label{obs1}
\en

\noi where $S_W$ is the standard Wilson action and $Z_W$ is given by
$\langle 1\rangle_W =1$. Given boundary conditions, the average
$\langle \cao\rangle_W$ is (mathematically) well defined and can be
calculated numerically.

In the TM approach the connection between the TM $\hatV$ and the
Hamiltonian $\hatH$ is given by $\hatV = e^{-a\hatH}$. The average of
any operator ${\widehat\cao}$ is

\eq
\langle{\widehat\cao}\rangle_H = \frac{1}{Z_H}\tr \left( \hatV^{N_4}
{\widehat\cao}P_0 \right)
            \label{obs2}
\en

\noi where $P_0$ projects on colorless states and $N_4$ is a number of
sites along the forth direction. The partition function $Z_H~$ is given
by $~\langle 1\rangle_H =1~$. The consistency between two approaches,
i.e. $~Z_W=Z_H \equiv Z~$, defines the transfer matrix $~\hatV~$
\cite{cre1,lues,cre2}.

Here it will be shown that the connection between operators
$\widehat\cao$ and corresponding functionals $~\cao(U;\psi;\opsi)~$ is
more complicated than in the continuum.

It will be also shown a connection between b.c.'s and the choice of the
fermionic Fock space (FFS). Another b.c.'s for $\psi,\opsi$ will be
proposed.

\section{Transfer matrix formalism}

Let $c_i^{\dagger}(c_i)$ and $d_i^{\dagger}(d_i)$  be
creation(annihilation) operators of quarks and antiquarks,
respectively,  $~[ c_i,c^{\dagger}_j]_{+} = [
d_i,d^{\dagger}_j]_{+} = \delta_{ij}~$ and $i,j=(\vx;\alpha;s)~$ where
$\vx~$ is a $3d$ coordinate, $\alpha~$ is a color index and $s~$ is a
spin index.

Let $~\{U_{\vx;k}\}~$ and $~\{U_{\vx;k}^{\prime}\}~$ be two gauge field
configurations defined on spacelike links. The transfer matrix $\hatV$
is an integral operator with respect to the gauge fields. Its kernal
$~V(U;U^{\prime})~$ is an operator in the fermion Hilbert space
\cite{lues} :

\eqa
V(U;U^{\prime}) &=& \hatT^{\dagger}_F(U)V_G(U;U^{\prime})
\hatT_F(U^{\prime})~;
          \label{transf_luescher}
\\
\hatT_F(U) &=& C_0\, e^{ d^TQ(U)c}
e^{ -c^{\dagger}R(U)c - d^{\dagger}R^T(U)d}
\nonumber
\ena

\noi where $~V_G(U;U^{\prime})~$ corresponds to a pure gauge part
\cite{cre1}. Matrix $Q$ is

\eq
Q_{\vx\vy} = \frac{i}{2}\sum_{k=1}^3\Bigl[ \delta_{\vy, \vx+k}
U_{xk} - \delta_{\vy,\vx-k}U_{x-k,k}^{\dagger} \Bigr] \sigma_k~.
\en

\noi and $R$ is given by $~e^R = \bfB^{1/2}/\sqrt{2\kappa}~$ with

\eq
\bfB_{\vx\vy} = \delta_{\vx\vy} - \kappa\sum_{k=1}^3\Bigl[
\delta_{\vy, \vx+k}U_{xk}+ \delta_{\vy,\vx-k}U_{y,k}^{\dagger} \Bigr]~.
\en

\noi $C_0$ is a constant, $\kappa$ is a hopping parameter and
$\sigma_k$ are Pauli matrices. One can prove the equality $~Z_H=Z_W~$
using the Grassmannian coherent state basis
$~|\eta\zeta\rangle = e^{ c^{\dagger}\eta
+ d^{\dagger}\zeta } |0\rangle~$ and $~\langle\oeta\ozeta|
= \langle 0| e^{ \oeta c + \ozeta d }~$,
where $~\eta_{\vx}(x_4),\ldots,\ozeta_{\vx}(x_4)~$ are
Grassmannian variables and $~c^{\dagger}\eta =\sum_{\vx}
c_{\vx}^{\dagger}\eta_{\vx}~$, etc.. The bispinors
$~\psi_{\vx}(x_4),\,\opsi_{\vx}(x_4)~$ are given by

\eq
\psi(x_4) = {\widetilde \bfB} \left(
\hspace{-1mm}
\begin{array}{c}
   \eta      \\
   \ozeta^T
\end{array}
\hspace{-1mm}
\right);
\quad
\opsi(x_4) = \left(
\hspace{-1mm}
\begin{array}{c}
   \oeta^T      \\
   \zeta
\end{array}
\hspace{-1mm}
\right)^T
\gamma_4 {\widetilde \bfB}
           \label{new_grass}
\en

\noi where $~{\widetilde \bfB}_{\vx\vy}=\bfB^{-1/2}_{\vx\vy}~$. Note
that on the lattice the connection between $~\psi,\opsi~$ and
$~\eta,\ldots,\ozeta~$ is rather nontrivial.

Another important point is the choice of the fermionic Fock space.
Assuming that the FFS spanned by all possible fermionic states

\eq
|\{n_i\};\{m_j\}\rangle = \prod_{i=1}^N (c_i^{\dagger} )^{n_i}
\prod_{j=1}^N (d_j^{\dagger} )^{m_j}|0\rangle~,
              \label{fock1}
\en

\noi where $~n_i,m_j=0,1~$, one arrives at b.c.'s

\eq
\psi_{\vx}(L_4) = -\psi_{\vx}(0)~;
\qquad
\opsi_{\vx}(L_4) = -\opsi_{\vx}(0)~
            \label{antiper}
\en

\noi and $~U_{\vx k}(L_4) = U_{\vx k}(0)~$ where $~L_4=N_4a$.

\section{Fermionic currents}

\subsection{Pseudoscalar current}

Pseudoscalar current $~\hatJ_{\vx}^{P}~$ is given by

\eq
\hatJ_{\vx}^{P} = ~ :\! i\chi^{\dagger}_{\vx}\gamma_4\gamma_5
\chi_{\vx}\! :
~=~ i(c^{\dagger}_{\vx}d^{\dagger\, T}_{\vx} - d_{\vx}^Tc_{\vx})~,
\en

\noi where $~\chi_{\vx} = \left(
\begin{array}{c}
   c_{\vx}                \\
   d^{\dagger\, T}_{\vx}  \\
\end{array} \right)~,
~~\chi_{\vx}^{\dagger} = ( c_{\vx}^{\dagger} ~~~d_{\vx}^T)
~$
and $\gamma_\nu$ are euclidian $~\gamma$--matrices. The average is

\eq
\langle \hatJ^{P}\rangle_H
\equiv \frac{1}{Z}\tr \left( \hatV^{N_4}\hatJ^{P} P_0 \right)
\en

\noi where $~\hatJ^{P} = \sum_{\vx} \hatJ_{\vx}^{P}~$. Following
\cite{lues}, let us choose the FFS as in eq.~(\ref{fock1}). One obtains
\cite{mitr}

\eq
\langle \hatJ^{P}\rangle_H = \frac{1}{Z}\int\![dU][d\opsi\psi]
~J^{P}(\psi,\opsi,U) \cdot e^{-S_W}
\en

\noi where Grassmannian current $J^{P}(\psi,\opsi,U)$ is

\eq
J^{P} = -i\sum_{\vx\vy} \opsi_{\vx}(0)\bfB_{\vx\vy}(0)\gamma_5
\psi_{\vy}(0)
\en

\noi and boundary conditions given in eq.~(\ref{antiper}).
The zero--momentum pseudoscalar correlator is

\eqa
\lefteqn{
\Gamma^{P}(\tau) = \frac{1}{Z}\tr \left( \hatV^{N_4-\tau}
\hatJ^{P} \hatV^{\tau}\hatJ^{P}P_0 \right)    }
\nonumber \\
&=& \frac{1}{Z}\int\![dU][d\opsi\psi]~J^{P}(\tau)
J^{P}(0) \cdot e^{-S_W}~.~~~
\ena

\noi Note that $J^{P}$ depends on fields $U_{x\mu}$ and does not
coincide with the naive expression

\eq
J^{P}_{naive} = (\opsi\gamma_5\psi)(0)
= \sum_{\vx} \opsi_\vx(0)\gamma_5\psi_\vx(0)~.
\en

\subsection{Scalar current}

Pseudoscalar current $~\hatJ_{\vx}^{S}~$ is given by

\eq
\hatJ^{S}_\vx \, =\, :\!\chi^{\dagger}_{\vx}\gamma_4\chi_{\vx}\! :
~=~ c^{\dagger}_\vx c_\vx + d^{\dagger}_\vx d_\vx~,
\en

\noi and $~\hatJ^{S} = \sum_{\vx} \hatJ_{\vx}^{S}~$.
One obtains  \cite{mitr}

\eqa
\lefteqn{
\langle \hatJ^{S}\rangle_H = \frac{1}{Z}\tr\left(\hatJ^{S}\,\hatV^{N_4}
P_0 \right)   }
\nonumber \\
&=& \frac{1}{Z} \int\![dU][d\opsi d\psi]~J^S(\psi,\opsi,U)\cdot e^{-S_W}
          \label{scalar_cond}
\ena

\noi where Grassmannian current $J^{S}(\psi,\opsi,U)$ is

\eqa
\lefteqn{
J^{S} = 2\kappa\Bigl[ \opsi(a)P^{(+)}_4 U^{\dagger}_4(0)\psi(0) + }
\\
&& \opsi(0)P^{(-)}_4 U_4(0)\psi(a)
- 2\opsi(0)P^{(-)}_4\bfC(0)\psi(0)\Bigr] .
\nonumber
\ena

\eq
\bfC_{\vx\vy}(x_4) = \frac{1}{2}\sum_{k=1}^3\Bigl[ \delta_{\vy,\vx+k}
U_{xk} - \delta_{\vy,\vx-k}U_{y,k}^{\dagger} \Bigr]\gamma_k
\en

\noi Evidently, $J^S$ does not coincide with

\eq
J^{S}_{naive} = (\opsi\psi )(0) =
\sum_{\vx}\opsi_{\vx}(0)\psi_{\vx}(0)~.
\en

\subsection{Non--zero chemical potential $~\mu$}

The partition function $Z(\mu)$ with nonzero chemical potential $~\mu~$
is given by

\eq
Z(\mu) = \tr \left( \hatV^{N_4} e^{ \mu\hatN/T}P_0\right)
\en

\noi where $~\hatN = \sum_\vx (c^{\dagger}_\vx c_\vx - d^{\dagger}_\vx
d_\vx)~$. One obtains

\eq
Z(\mu) = \int\![dU][d\opsi d\psi]
~\exp\left\{ -S_W + \delta S_F\right\}~,
       \label{Z_lambda_bar}
\en

\noi where

\eqa
\lefteqn{
\delta S_F = 2\kappa \sum_{\vx}\Bigl[ (e^{\mu/T}-1)\opsi_{\vx}(a)
P^{(+)}_4 U^{\dagger}_{\vx 4}(0)\psi_{\vx}(0)  }
\nonumber \\
&& + (e^{-\mu/T}-1)\opsi_{\vx}(0)P^{(-)}_4 U_{\vx 4}(0)\psi_{\vx}(a)
\Bigr].
\ena

\noi Making the change of variables

\eqa
\psi_{\vx}(x_4) &\to&  \left\{
\begin{array}{rl}
  e^{ -x_4\mu} \psi_{\vx}(x_4) & ~x_4\ne 0   \\
  e^{ -L_4\mu} \psi_{\vx}(x_4) & ~x_4 = 0
\end{array} \right.
\\
\opsi_{\vx}(x_4) &\to&  \left\{
\begin{array}{rl}
  e^{ x_4\mu} \opsi_{\vx}(x_4) & ~x_4\ne 0   \\
  e^{ L_4\mu} \opsi_{\vx}(x_4) & ~x_4 = 0
\end{array} \right. , ~~~
\ena

\noi one obtains \cite{mitr}  (see also \cite{palu}) the modified
fermionic matrix $~\cam(U;\mu)= {\hat 1} -2\kappa D(U;\mu)~$  where

\eqa
\lefteqn{
D_{xy} = \sum_{k=1}^3\Bigl[ \delta_{y, x+k}P^{(-)}_k U_{xk}
+ \delta_{y,x-k}P^{(+)}_k U_{x-k,k}^{\dagger} \Bigr]    }
\nonumber \\
&& \hspace{-5mm} +
\Bigl[ e^{-a\mu}\delta_{y;x+{\hat 4}}P^{(-)}_4 U_{x4} + e^{a\mu}
\delta_{y;x-{\hat 4}} P^{(+)}_4 U_{x-{\hat 4};4}^{\dagger} \Bigr].
\nonumber
\ena

\noi Evidently, $~\cam(U;\mu)~$ coinsides with the fermionic matrix for
the non--zero chemical potential proposed many years ago in \cite{haka}.

\section{Fermionic Fock space and boundary conditions}

The important observation is that b.c.'s for Grassmannian variables
$\psi,\opsi$ along the forth direction depend on the choice of the FFS.
This choice depends on the model (physical) assumptions.

For example, QCD vacuum is supposed to have an equal number of quarks
and antiquarks, and in the zero temperature limit vacuum eigenstate is
expected to give a main contribution. So, one can choose the fermionic
Fock (sub)space as in eq.(\ref{fock1}) but with additional condition

\eq
\sum_i n_i = \sum_i m_i~.
\en

\noi In the infinite volume limit one obtains \cite{mitr}

\eq
Z = \int_0^{2\pi}\frac{ d\varphi}{2\pi}\int\! [dU]
\int\![d\psi d\opsi]~e^{-S_W(U;\psi;\opsi)}
\en

\noi with fermionic boundary conditions

\eq
\psi_{\vx}(L_4) = -e^{i\varphi}\psi_{\vx}(0) ;
~\opsi_{\vx}(L_4) = -e^{-i\varphi}\opsi_{\vx}(0)
            \label{bound}
\en

\noi One may expect that at $N_4<\infty$ these b.c.'s could be a better
choice for the zero temperature calculations, e.g., for the hadron
spectroscopy study.

Another interesting case is a finite temperature transition in the
(early) Universe with zero baryon asymmetry : $~\Delta B=0~$. Note that
in this case for Polyakov loop $\capbig$ one obtains $~\langle
\capbig\rangle_W = 0~$ and $\langle |\capbig|\rangle_W $ is expected to
be a good order parameter as in quenched QCD.

\section{Summary}

A connection between operator current $\hatJ$ and corresponding
Grassmannian current $J$ is shown to be more complicated than in
continuum. In particular, $J$'s depend on fields $~U_{x\mu}~$, i.e.
$J=J(\psi,\opsi,U)~$, and $J\ne J_{naive}$.

The modified action with non--zero chemical potential $~\mu\ne 0$ is
derived.

A choice of the  b.c.'s along the forth direction and their connection
to FFS is discussed. Another b.c.'s for $\psi,\opsi$ are proposed which
could be a better choice for, e.g, the hadron spectroscopy study.

\end{document}